\documentclass[10pt,prl,twocolumn,amsmath,amssymb,superscriptaddress,longbibliography,aps]{revtex4-2}
\usepackage[T1]{fontenc}
\usepackage{graphicx}
\usepackage{xcolor}
\usepackage[english]{babel}
\usepackage{hyperref}

\begin{document}

\title{Single-site diagonal quantities capture off-diagonal long-range order}

\author{M. Sanino}
\affiliation{S\~ao Paulo State University (UNESP), Institute of Chemistry, 14800-090, Araraquara, S\~{a}o Paulo, Brazil.}

\author{I. D\textquoteright Amico}
\affiliation{School of Physics, Engineering and Technology, University of York, York YO10 5DD, United Kingdom.}

\author{V. V. Fran\c{c}a}
\affiliation{S\~ao Paulo State University (UNESP), Institute of Chemistry, 14800-090, Araraquara, S\~{a}o Paulo, Brazil.}

\author{I. M. Carvalho}
\affiliation{S\~ao Paulo State University (UNESP), Institute of Chemistry, 14800-090, Araraquara, S\~{a}o Paulo, Brazil.}

\date{\today}

\begin{abstract}

Quantum phase transitions are typically marked by changes in quantum correlations across various spatial scales within the system. A key challenge lies in the fact that experimental probes are generally restricted to diagonal quantities at the single-site scale, which are widely believed to be insufficient for detecting phases with off-diagonal long-range order, such as superconducting states. In a striking departure from conventional expectations, we show that single-site diagonal descriptors --- charge and spin fluctuations, occupation probabilities, and entanglement --- can capture the emergence of off-diagonal long-range order in the one-dimensional extended Hubbard model at half-filling. These single-site quantities display clear critical signatures of the superconducting transition, preceded by a continuous breaking of particle-hole symmetry, consistent with a second-order phase transition. While this symmetry breaking has a negligible effect on single-site descriptors, it allows a direct connection between local fluctuations and nonlocal correlations.

\end{abstract}

\maketitle


Precise control of quantum systems is now a common feature across various quantum simulators, such as trapped ions \cite{Richerme2014,doi:10.1126/sciadv.1700672}, ultracold atoms \cite{Kinoshita2006,doi:10.1126/science.aaa7432}, nitrogen-vacancy centers \cite{Choi2017}, and Rydberg atoms \cite{Bernien2017,Browaeys2020}. These experiments not only provide a highly tunable platform for engineering and exploring strongly interacting systems \cite{PhysRevLett.111.185305,PRXQuantum.3.020303,Yan2013,doi:10.1126/science.aav9105,doi:10.1126/science.aat4134,doi:10.1126/science.aag3349,doi:10.1126/science.aag1635,doi:10.1126/science.aam7838}, but also grant access to observables at the level of individual sites \cite{Christakis2023,Gross2021,PhysRevLett.103.080404}, making it possible to directly explore local fluctuations and nonlocal correlations in real space \cite{doi:10.1126/science.aam8990,Koepsell2019,Salomon2019}. 

Such advances in single-site probing open new avenues for exploring strongly interacting systems, where transitions without long-range order have been successfully captured via single-site diagonal descriptors. Examples include spin and charge density waves \cite{PhysRevLett.93.086402,PhysRevB.79.245130,PhysRevResearch.6.L022064}, which exhibit diagonal order; off-diagonal metallic phases \cite{Karlsson_93_2011,PhysRevA.85.033612,PhysRevLett.100.070403,canellaVV}; and superfluid and topological phases emerging in confined systems \cite{Feguin_76_2007,Liao_467_2010,Canella2020,Sanino_2_2025}.
Nevertheless, some of the most intriguing phases --- such as high-$T_C$ superconductivity and bond-ordered waves \cite{PhysRevLett.125.017001,PhysRevLett.116.225305,PhysRevLett.106.136402,Qu2022,EjimaEHM2007,PhysRevLett.129.076403} --- exhibit off-diagonal long-range order (ODLRO), rooted in phase coherence and entanglement across distant sites. It is widely assumed that such nonlocal features can only be accessed through off-diagonal observables, under the intuitive expectation that single-site diagonal descriptors are insensitive to phase coherence. This view implies a fundamental limitation in capturing transitions into ODLRO phases using only diagonal quantities.

\begin{figure}[htbp]
\begin{centering}
\includegraphics[scale=0.45]{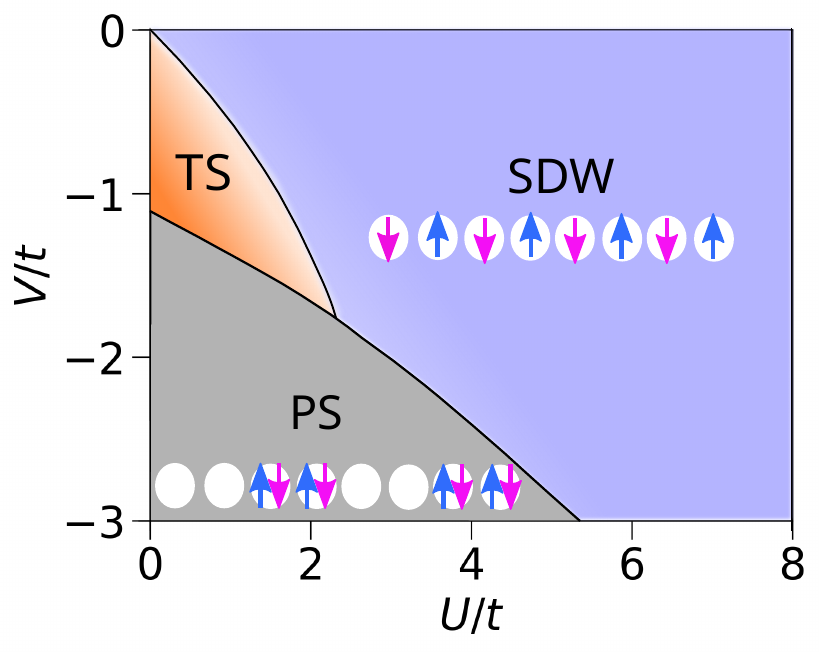}
\par\end{centering}
\caption{(Color online) Phase diagram of the 1D extended Hubbard model at half-filling. It includes two insulating phases: the spin density wave (SDW), characterized by a degenerate ground state with electrons alternating between spin states $\sigma = \uparrow, \downarrow$; and a highly degenerate phase separation (PS), identified by doubly occupied clusters. A triplet superconducting (TS) phase lies in between the insulating regimes for small $U/t$. Adapted from \cite{Qu2022}.} \label{fig1}
\end{figure}

A typical example in this context is the detection of symmetry-breaking transitions within the one-dimensional extended Hubbard model (EHM) --- a well-established framework for describing strongly correlated systems \cite{essler2005one}, recently employed to model the cuprate compound ${\rm Ba}_{2-x}{\rm Sr}_{x}{\rm Cu}{\rm O}_{3+\delta}$ \cite{Wang_PRL2021,Qu2022}. For the half-filled EHM with on-site repulsion ($U>0$) and nearest-neighbor attraction ($V<0$), transitions into the superconducting phase have been successfully identified using off-diagonal descriptors \cite{Qu2022,PhysRevB.92.075423,PhysRevB.105.115145}. These developments have led to a detailed mapping of the phase diagram (see Fig.~\ref{fig1}), notably predicting the emergence of a gapless $p$-wave superconducting phase in one-dimensional cuprate chains.
In contrast, previous studies have shown that measures based solely on diagonal information fail to detect transitions associated with ODLRO \cite{PhysRevLett.93.086402,PhysRevB.74.045103,PhysRevB.79.245130}. Nevertheless, those works probed transitions through pronounced features in observables --- such as sharp peaks or discontinuities --- while setting aside subtler signatures that could be associated with finite-size effects.
Although recent complexity-based analyses of diagonal quantities have achieved some success in signaling transitions associated with ODLRO \cite{doi:10.1073/pnas.2004976117,PhysRevResearch.6.L022064,PhysRevA.109.053304}, they remain limited to the repulsive regime of the EHM and do not explicitly extract the underlying mechanisms or clarify the limitations that prevented their identification in earlier approaches.

In this work, by investigating ODLRO superconducting phases in the EHM, we show that several single-site diagonal descriptors --- spin and charge fluctuations, occupancy probabilities,  and entanglement entropy --- are capable of detecting the superconducting transition, exhibiting a clear extremum at the critical point that is not attributable to finite-size effects. Remarkably, we find that the transition is preceded by the continuous breakdown of particle-hole symmetry, a distinctive feature that reveals the underlying mechanism by which diagonal descriptors capture the critical nature of the transition. Even more intriguingly, this symmetry breaking has a negligible impact on single-site observables, allowing for a direct connection between local charge and spin fluctuations and nonlocal quantum entanglement.
This result not only enables probing nonlocal correlations through local fluctuations but also reveals how charge and spin degrees of freedom interconnect locally.

The Hamiltonian of the 1D EHM considered here, including nearest-neighbor electron hopping ($t$), on-site ($U$) and nearest-neighbor ($V$) interactions, is given by
\begin{eqnarray}
H&=&-t\sum_{j=1,\sigma}^{L-1}\left(a_{j,\sigma}^{\dagger}a_{j+1,\sigma}+a_{j+1,\sigma}^{\dagger}a_{j,\sigma}\right) \nonumber \\&+&U\sum_{j=1}^{L}n_{j\uparrow}n_{j\downarrow}+V\sum_{j=1}^{L-1}n_{j}n_{j+1},
\label{eq_EHM}
\end{eqnarray}
where $L$ is the lattice size, $a_{j\sigma}^{\left(\dagger\right)}$ is the electron annihilation (creation) operator, $\sigma=\uparrow,\downarrow$ labels the electron spin, and $n_{j}=n_{j\uparrow}+n_{j\downarrow}$ is the particle number operator at site $j$. We focus on the regime of on-site repulsion ($U>0$) and  attractive nearest-neighbor interaction  ($V<0$), consistent with the parameter space relevant to cuprate chains \cite{ZXShen_Science2021,Wang_PRL2021,Qu2022,Kennedy_111_2025}. Throughout this study, we consider the half-filled case, regime at which the ground-state phase diagram, Fig. \ref{fig1}, is well-established in the literature \cite{Qu2022,PhysRevB.33.8155,nakamura2000tricritical}.
It features two distinct insulating phases, spin density wave (SDW) and phase separation (PS), and a spin-triplet superconducting phase (TS). These transitions were characterized using large-scale density matrix renormalization group (DMRG) simulations by computing off-diagonal descriptors, as the Luttinger parameters $K_\rho$, central charge $c$, and both the charge and spin structure factors \cite{Qu2022}. We use this phase diagram as a guiding map to analyze --- through single-site diagonal descriptors --- the SDW-TS and TS-PS transitions.

In order to determine the diagonal quantities at site $j$, we compute the associated reduced density matrix, $\rho_{j}={\rm Tr}_{L\neq j}\left|\varPsi\right\rangle \!\!\left\langle \varPsi\right|$, where $\left|\varPsi\right\rangle$ is the ground state and ${\rm Tr}_{L\neq j}$ denotes the partial trace over all sites except $j$. Since the Hamiltonian (\ref{eq_EHM}) conserves both total particle number and the $z$-component of spin, $\rho_j$ is diagonal in the occupation basis \cite{PhysRevB.79.245130} 
\begin{eqnarray}
    \rho_{j}=w_{\uparrow}\left|\uparrow\right\rangle \!\!\left\langle \uparrow\right|+w_{\downarrow}\left|\downarrow\right\rangle \!\!\left\langle \downarrow\right| +w_{0}\left|0\right\rangle \!\!\left\langle 0\right|+w_{2}\left|2\right\rangle \!\!\left\langle 2\right|.
    \label{reduced_density}
\end{eqnarray}
Here $w's$ denote the probabilities of finding, at site $j$, one of the four possible states: double occupancy with opposite-spin electrons, $\left|2\right\rangle$ ($w_2$); single occupancy with spin up $\left|\uparrow\right\rangle$ ($w_\uparrow$) or spin down $\left|\downarrow\right\rangle$ ($w_\downarrow$); or the empty (hole) state, $\left|0\right\rangle$ ($w_0$):
\begin{eqnarray}
    w_{2}&=&\left\langle n_{j\uparrow}n_{j\downarrow}\right\rangle,\nonumber \\w_{\uparrow}&=&\left\langle n_{j\uparrow}\right\rangle -w_{2} ,\nonumber \\ \nonumber w_{\downarrow}&=&\left\langle n_{j\downarrow}\right\rangle -w_{2}, \\  w_{0}&=&1-w_{\uparrow}-w_{\downarrow}-w_{2},
    \label{relations_w}
\end{eqnarray}
where the normalization condition $\sum_{k}w_{k}=1$ is satisfied. Notice that any local observable $\mathcal{L}_{j}$ can be entirely determined via Eq.(\ref{relations_w}) through $\left\langle \mathcal{L}_{j}\right\rangle ={\rm Tr}(\mathcal{L}_{j}\rho_{j})$\cite{cohen2019quantum,Davidson1976}. This includes not only the local density $\left\langle n_{j}\right\rangle$  and magnetization $\left\langle s_{j}^{z}\right\rangle=(\left\langle n_{j\uparrow}\right\rangle -\left\langle n_{j\downarrow}\right\rangle )/2$, but also their respective fluctuations (or variances), such as the charge fluctuation $\left\langle \delta n_j\right\rangle = \left\langle n_{j}^{2}\right\rangle -\left\langle n_{j}\right\rangle ^{2}$, and spin fluctuation $\ensuremath{\left\langle \delta s_j^{z}\right\rangle =\left\langle (s_{j}^{z})^2\right\rangle -\left\langle s_{j}^{z}\right\rangle ^{2}}$. In the zero-magnetization sector (and without spontaneous spin polarization), $\langle n_{j\uparrow} \rangle = \langle n_{j\downarrow} \rangle$, thus  $w_\uparrow = w_\downarrow$ throughout our analysis. 

Using (\ref{relations_w}) and the fermionic anti-commutation relations, one derives the following expressions for charge and spin fluctuations of a given site
\begin{eqnarray}
\left\langle \delta n\right\rangle &=& w_{0} + w_{2} - \left(w_{0} - w_{2}\right)^{2},\\
\label{charge_fluc}
\left\langle \delta s^{z}\right\rangle &=& w_{\uparrow}/2 = w_{\downarrow}/2.  
\label{spin_fluc}
\end{eqnarray}
Note that the term $(w_0 - w_2)^{2}$ introduces a quadratic (rather than linear) penalty for deviations from $w_0 = w_2$, reflecting the particle-hole symmetry. This symmetry corresponds to the transformation $a_{j,\sigma} \rightarrow (-1)^j a_{j,\sigma}^\dagger$, under which $n_{j,\sigma} \rightarrow 1-n_{j,\sigma}$, effectively exchanging empty and doubly occupied sites, i.e., $w_0 \leftrightarrow w_2$. 

The probabilities in (\ref{relations_w}) also allow for a direct calculation of the single-site entanglement entropy \cite{PhysRevLett.93.086402}
\begin{eqnarray}
    S=&-&w_{\uparrow}\log_{2}w_{\uparrow}-w_{\downarrow}\log_{2}w_{\downarrow} \label{von_Neumann} \\ \nonumber
    &-&w_{0}\log_{2}w_{0}-w_{2}\log_{2}w_{2} ,
\end{eqnarray}
which quantifies the entanglement between a given site and the rest of the system, and is known to signal quantum phase transitions through extrema or singularities as a function of a control parameter \cite{PhysRevLett.95.196406,RevModPhys.80.517,PhysRevLett.100.070403,Zawadzki_7_2024,Paulleti_644_2024,Vivian_77_2008,PhysRevB.101.214522,Canella_9_2019}.

To obtain the ground-state wavefunction of the Hamiltonian (\ref{eq_EHM}), we employ the DMRG method~\cite{dmrg2}, implemented using the ITensor library \cite{itensor}. Calculations are performed with open boundary conditions on systems of size up to $L=256$ at half-filling. A pinning field is applied at the edge sites to improve the convergence of the ground-state energy to at least $10^{-7}$. All our numerical results are taken at the central site, $j = L/2$, to avoid boundary effects on the single-site measures.

\begin{figure}[htbp]
\begin{centering}
\includegraphics[scale=0.4]{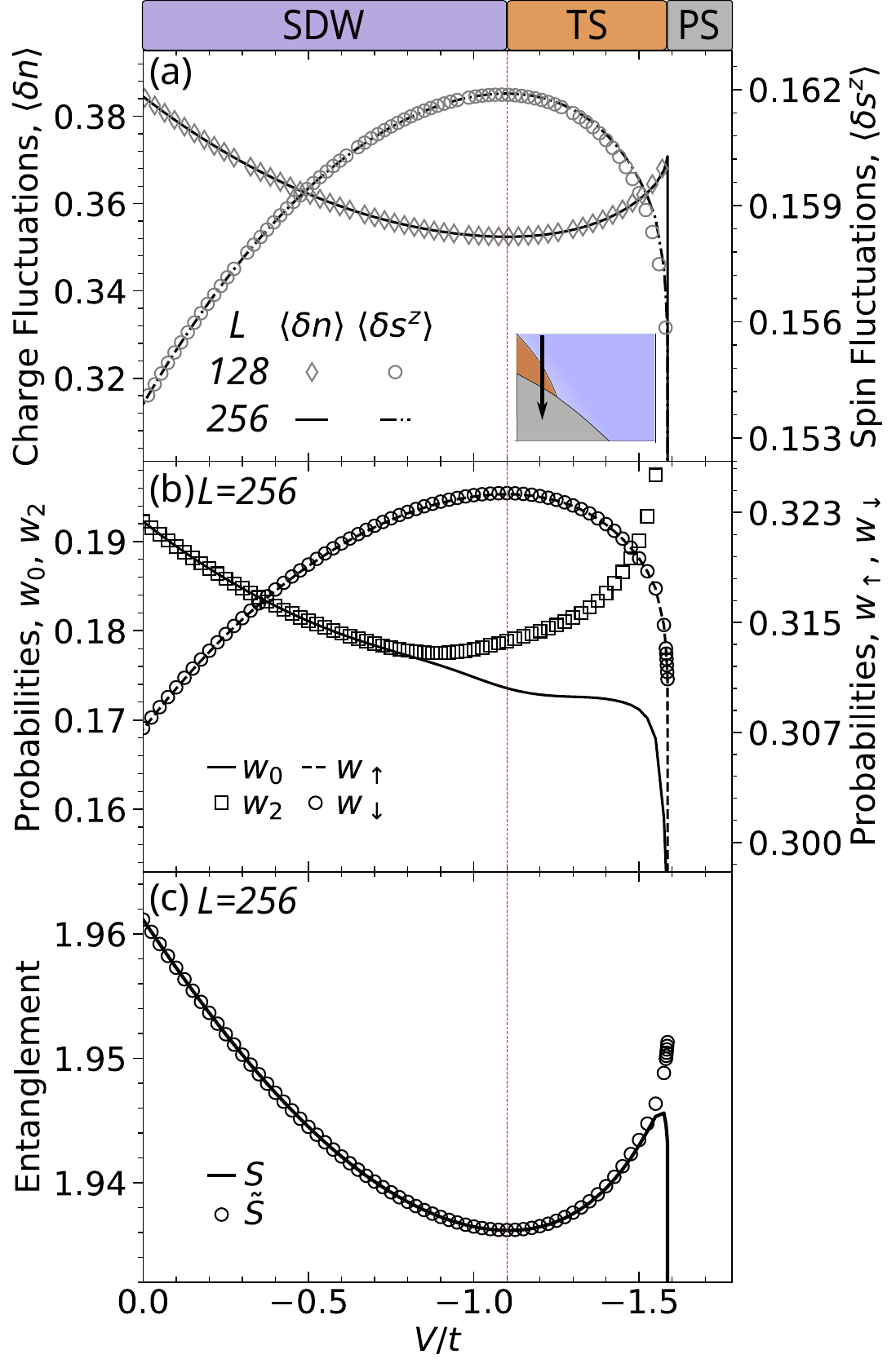}
\par\end{centering}
\caption{(Color online) Diagonal descriptors at the central site ($L/2$) under attractive nearest-neighbor interaction $V$, with a fixed onsite repulsion of $U = 1.6t$. The vertical dashed red line at $V_C = -1.1t$ marks the extremum associated with the critical transition to the triplet superconducting (TS) phase. At the first-order phase separation (PS) transition, occurring at $V = -1.55t$, all quantities exhibit an abrupt drop. The y-axis is rescaled to highlight the superconducting transition.} \label{fig2}
\end{figure}

We begin by presenting in Fig. \ref{fig2}(a) numerical results for charge and spin fluctuations as a function of the attractive nearest-neighbor interaction $V$, for a fixed onsite repulsion $U=1.6t$. It is remarkable that at $V_C = -1.1t$, both fluctuations exhibit extrema that not only signal the SDW-TS transition, but also remain unchanged when the system size is doubled from $L=128$. This robustness strongly indicates that the behavior persists in the thermodynamic limit, establishing it as a genuine signature of the transition. By further increasing $|V|$, an abrupt drop of both fluctuations is observed at $V \simeq -1.55t$, marking the TS-PS transition. Thus the TS phase domain identified here {\it via local diagonal descriptors}, $-1.55t < V \leq -1.1t$, closely matches that obtained from the decay of long-range off-diagonal TS correlations \cite{Qu2022}, $-1.55t \lesssim V \lesssim -1t$. An intriguing feature is the inverse relationship between charge and spin fluctuations observed across the SDW and TS regimes --- a pattern that will be clarified later.

Examining the occupation probabilities in Fig. \ref{fig2}(b), we find that 
the unpaired components $w_\uparrow=w_\downarrow$ closely track the spin fluctuations, in line with Eq.(\ref{spin_fluc}), while $w_2$ and $w_0$ exhibit a gradual breakdown of their symmetry ($w_2\sim w_0$) near the SDW-TS transition. This last feature clearly confirm the critical nature of the SDW-TS transition within the single-site perspective, in sharp contrast to the first-order TS-PS transition, where no symmetry breaking takes place and there is the sharp suppression of the remaining probabilities. Interestingly, the incipient local symmetry breaking preceding the critical point affects neither local fluctuations nor entanglement, as shown in the upper and lower panels of Fig.~\ref{fig2}.

Clarifying the insensitivity of local quantities to symmetry breaking involves examining charge fluctuations in Eq.~(\ref{charge_fluc}),  which penalizes it via the term $(w_0 - w_2)^2$. Across the SDW–TS transition, this contribution remains $\mathcal{O}(10^{-4})$, while $w_0 + w_2 = \mathcal{O}(10^{-1})$, indicating a negligible role in the behavior of charge fluctuations. Furthermore, by neglecting the contribution from $(w_0 - w_2)^2$ and employing the normalization condition $\sum_k w_k = 1$ in Eqs. (\ref{charge_fluc}) and (\ref{von_Neumann}), we obtain two key relations throughout the SDW and TS regimes. First, a direct relation between charge and spin fluctuations, given by  
\begin{eqnarray}
   \langle \delta n \rangle = 1-4\langle \delta s^{z} \rangle,
   \label{charge_spin_inter}
\end{eqnarray}
which captures the inverse relation between  $\langle \delta n \rangle$ and $\langle \delta s^{z} \rangle$, as observed numerically in Fig. \ref{fig2}(a). More interestingly, an approximation $\Tilde S$ to the single-site entanglement entropy $S$ as a function of the charge fluctuations, given by 
\begin{eqnarray}
    S\approx \Tilde{S}&=& 1 - \langle \delta n \rangle 
- \left(1 - \langle \delta n \rangle\right) \log_2\left[1 - \langle \delta n \rangle\right]\nonumber \\ 
&&- \langle \delta n \rangle \log_2 \left[ \frac{\langle \delta n \rangle}{2} \right].
    \label{von_Neumann_approximate}
\end{eqnarray}
This relation intrinsically links local fluctuations to nonlocal correlations. Notably, Eqs. (\ref{charge_spin_inter}) and (\ref{von_Neumann_approximate}) hold universally in regimes where the term $(w_0 - w_2)^2$ becomes negligible, highlighting their broad applicability.

Fig. \ref{fig2}(c) shows both $S$ and $\Tilde{S}$, which remain nearly identical throughout the SDW and TS phases, confirming that the breaking of particle-hole symmetry has little impact on entanglement within these regimes. Both measures exhibit a local minimum at $V_C = -1.1t$, signaling the ODLRO second-order SDW-TS transition, followed by a sharp drop at $V = -1.55t$, marking the first-order TS-PS transition. Thus all the single-site quantities exhibit a sharp discontinuity at the TS-PS transition point $V = -1.55t$: charge and spin fluctuations and entanglement abruptly drop to zero, while the site becomes fully occupied ($w_2 = 1$). In contrast, $\Tilde{S}\rightarrow 1$, as $\langle \delta n \rangle\rightarrow 0$, and the approximation to the entanglement breaks down since the condition $w_0 + w_2 \gg (w_0 - w_2)^2$ no longer holds. Thus, the markedly different behaviors of diagonal descriptors across the two transitions help elucidate why the SDW–TS critical point was missed in previous analyses by \cite{PhysRevLett.93.086402, PhysRevB.74.045103}, which focused on sharp features.

In summary, our work fills a crucial gap by demonstrating that transitions involving off-diagonal long-range order can indeed be detected using bare diagonal descriptors. This advances the understanding of phase transitions through single-site measures and highlights the intricate link between charge and spin fluctuations and their connection to nonlocal correlations. Exemplified by the ODLRO superfluid SDW-TS transition in the 1D extended Hubbard model at half-filling, we show that all the analyzed single-site quantities --- charge and spin fluctuations, occupation probabilities and entanglement --- display extrema at the transition point, which persist in the thermodynamic limit. Our findings also reveal that the transition is preceded by a continuous breaking of particle-hole symmetry, which has minimal impact on fluctuations and single-site entanglement within SDW and TS phases. General relations were derived (i) explaining the inverse correlation between charge and spin fluctuations and (ii) intrinsically connecting local charge fluctuations to nonlocal correlations. 

We would like to thank M. F. Cavalcante for the fruitful discussion in the early stages of this study. This research was supported by the S\~ao Paulo Research Foundation (FAPESP), Brasil (2021/06744-8; 2023/00510-0; 2023/02293-7) and by CNPq (403890/2021-7; 306301/2022-9).


\sloppy
\bibliographystyle{apsrev4-2}
%


\end{document}